\documentclass[twocolumn,aps,pre]{revtex4}
\usepackage{graphicx,color}
\newcommand{\outersoln}{%
\begin{figure}[htbp]
   \includegraphics[width=3in,clip]{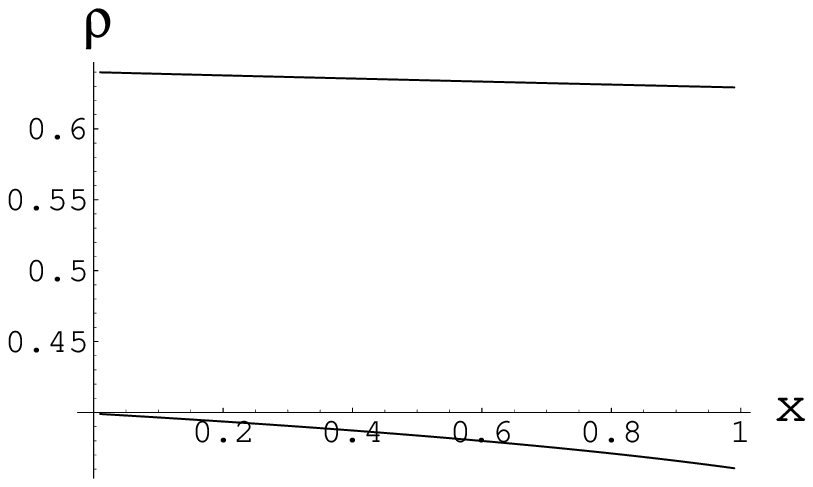}
    \caption{The plot of the left (upper curve) and 
right (lower curve) outer solutions. The left outer
solution satisfies the left boundary condition $\alpha=.64$. The right 
outer solution  satisfies the right boundary condition $\gamma=.36$. 
We have plotted the solutions till the other end of the lattice.}
\label{fig:calloutersoln}
\end{figure}
}
\newcommand{\currentdensity}{%
\begin{figure}[htbp]
   \includegraphics[width=3in,clip]{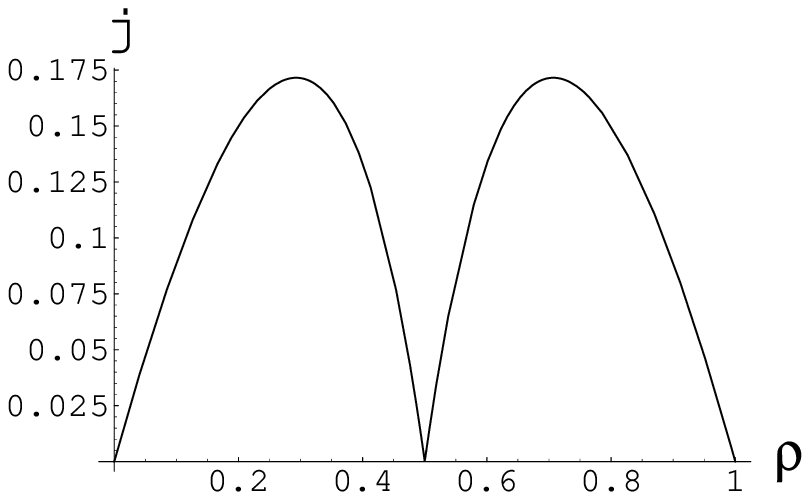}
    \caption{The plot of the particle current versus particle density 
for $\delta=0$ and $\Delta=1$.}
\label{fig:callcurrentplot}
\end{figure}
}
\newcommand{\picturedownshock}{%
\begin{figure}[htbp]
   \includegraphics[width=2in,clip]{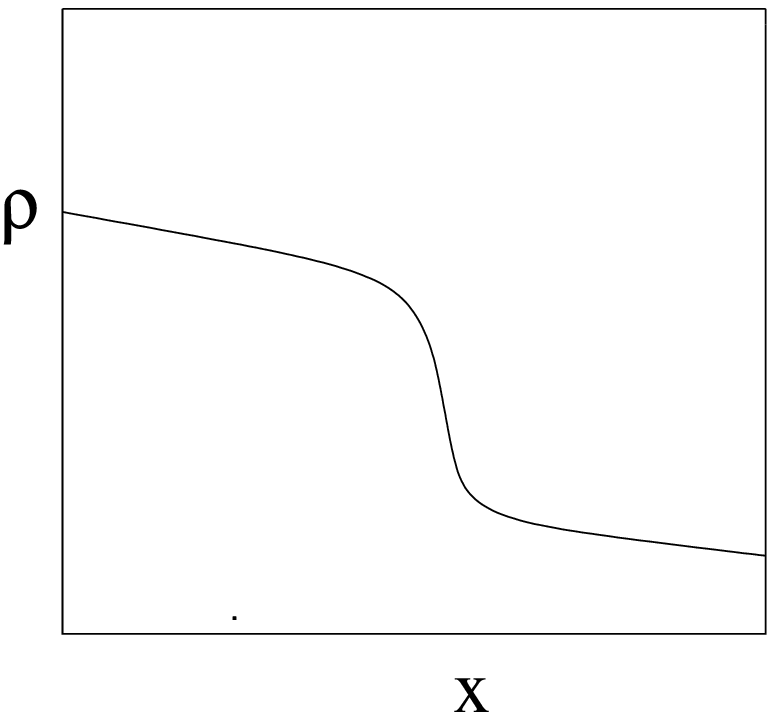}
    \caption{A diagram for the density profile with a downward shcok.}
\label{fig:downshock}
\end{figure}
}
\newcommand{\translines}{%
\begin{figure}
\centering
   \includegraphics[width=3.5in,clip]{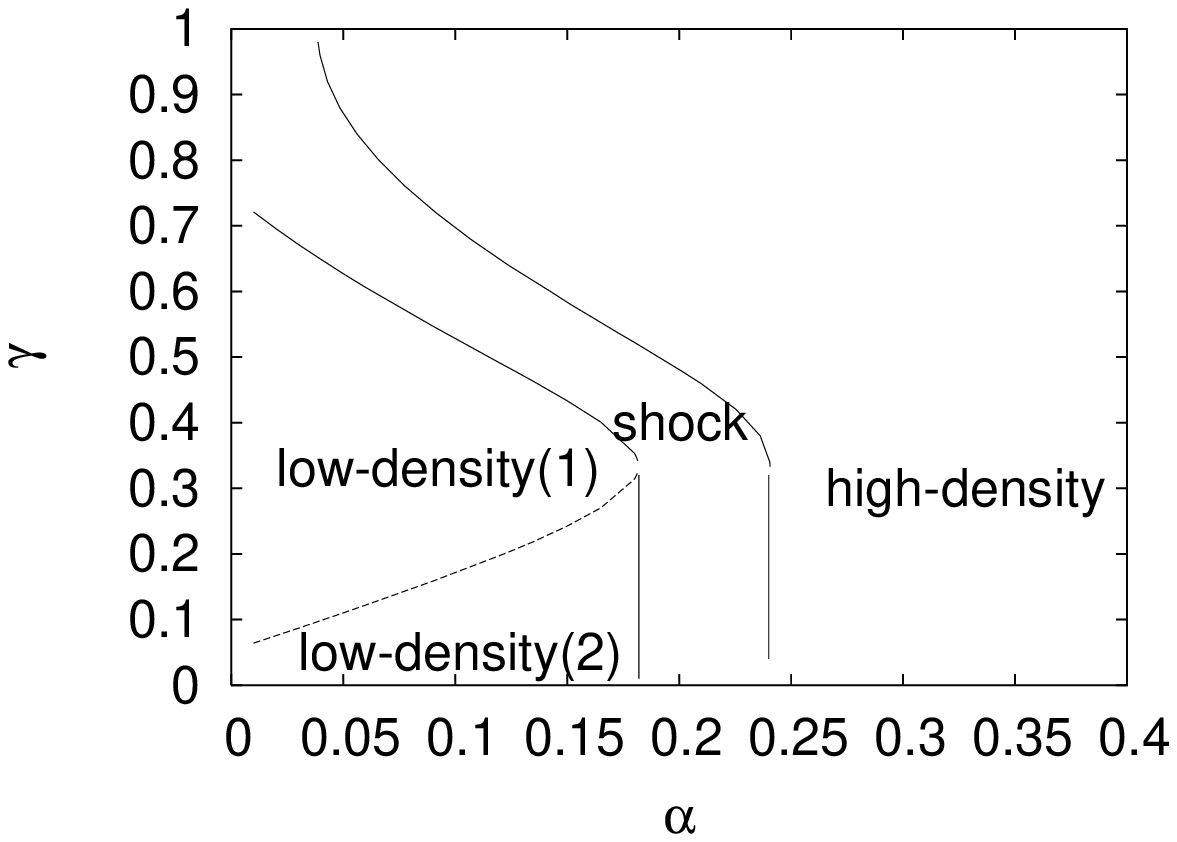}
    \caption{ Quantitative plot of the phase diagram  
for $\delta=1$ and $\Delta=0$ and $\Omega=.1$. Low-density (1) 
and low-density (2) are the two low-density phases in which the 
bulk profiles remain same but the surface layers have
 different slopes. These two phases are thus separated by the boundary
transition line. The phase boundary between the 
low-density and shock phases  and the boundary 
transition line  
 meet at the critical point $\{\alpha=.183..,\gamma=1/3\}$.}
\label{fig:phasediag}
\end{figure}
}
\newcommand{\highdensityplot}{%
\begin{figure}[htbp]
   \includegraphics[width=3in,clip]{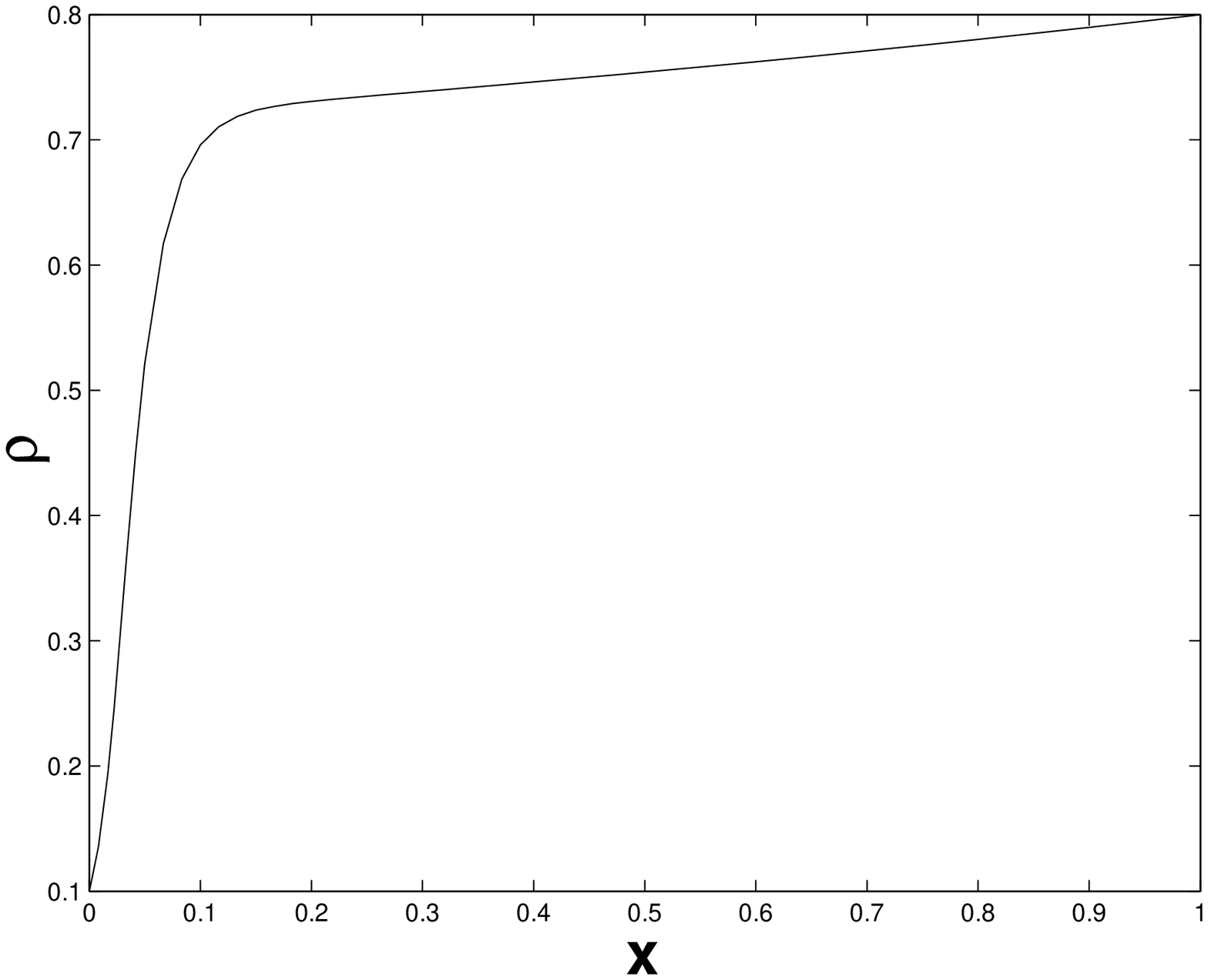}
    \caption{Plot of the density profile  for 
$\alpha=.1$, $\gamma=.8$, $\Omega=.1$ and $\epsilon=.034$. At these values of 
parameters, the system is in the high-density phase.}
\label{fig:highdensity}
\end{figure}
}
\newcommand{\shocktrans}{%
\begin{figure}[htbp]
   \includegraphics[width=3in,clip]{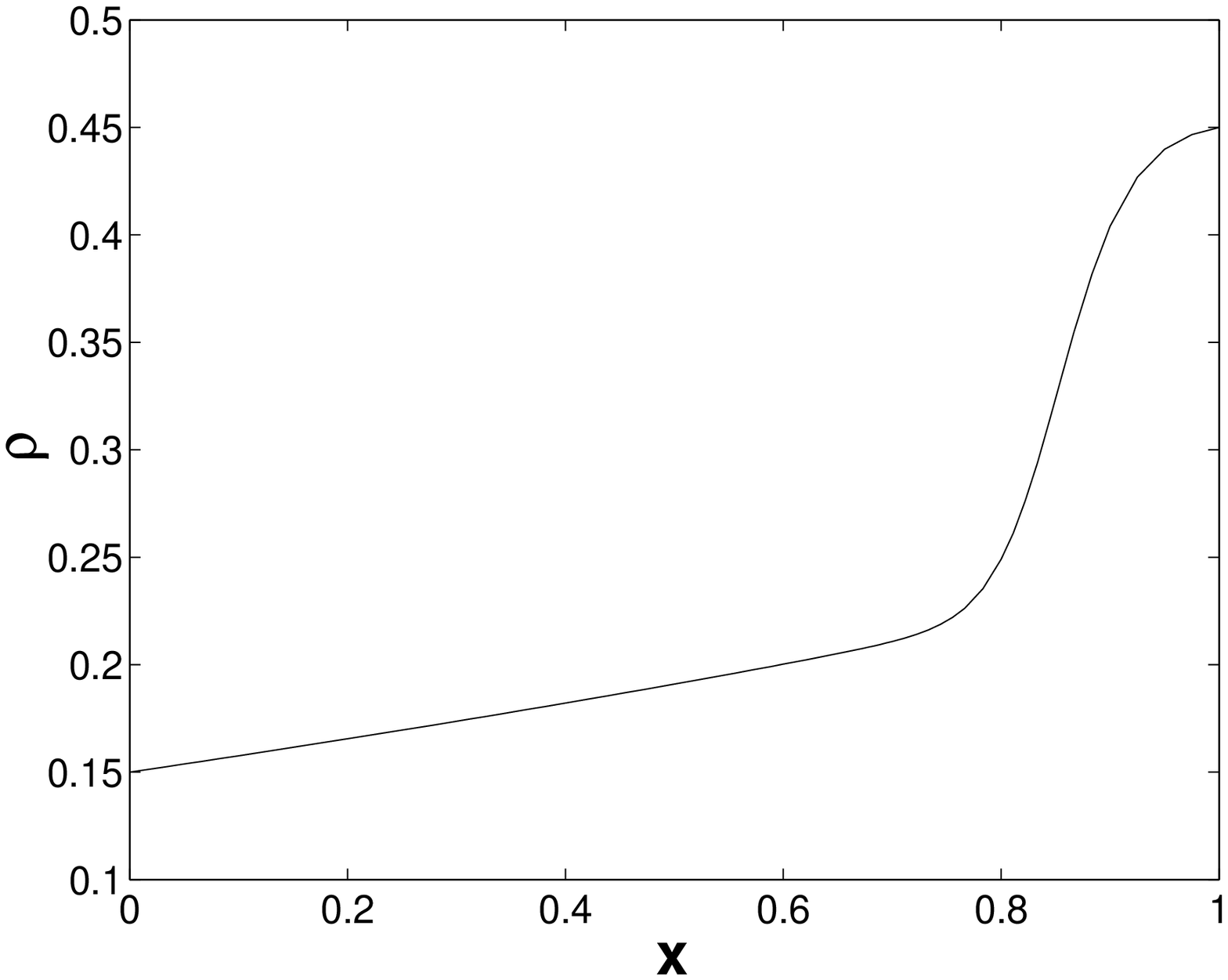}
    \caption{ Plot of the density profile for $\alpha=.15$, $\gamma=.45$,
$\Omega=.1$, and $\epsilon=.03$. At these values of the parameters, the 
boundary layer is just deconfined from the $x=1$ boundary.}
\label{fig:justdeconfine}
\end{figure}
}
\newcommand{\lowdensity}{%
\begin{figure}[htbp]
   \includegraphics[width=3in,clip]{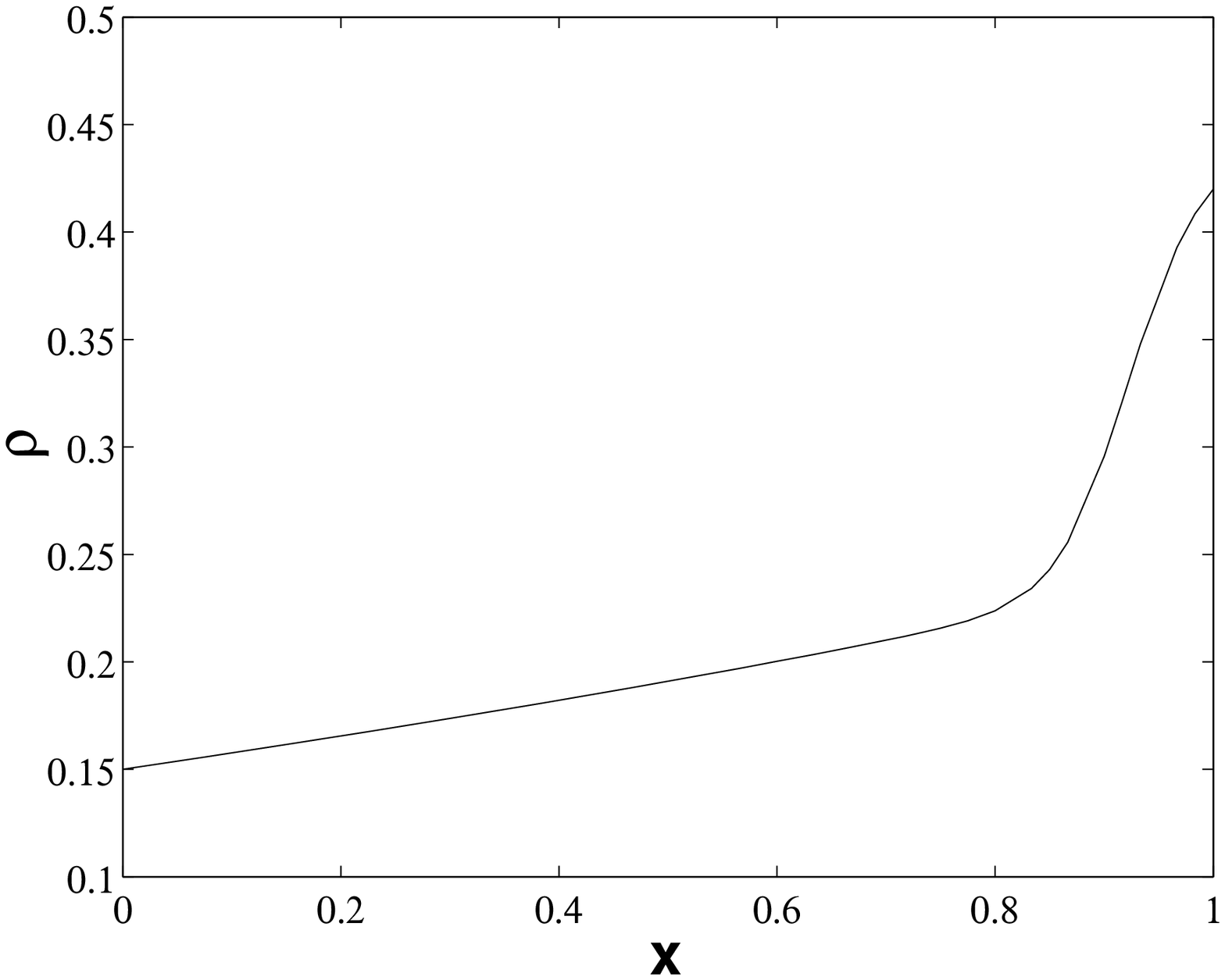}
    \caption{ Plot of the density profile for $\alpha=.15$, $\gamma=.42$,
$\Omega=.1$, $\epsilon=.03$. At these values of the parameters, the 
system is in the low-density phase, just below shockening transition line.}
\label{fig:justlowdensity}
\end{figure}
}

\newcommand{\differshock}{%
\begin{figure}[htbp]
   \includegraphics[width=3in,clip]{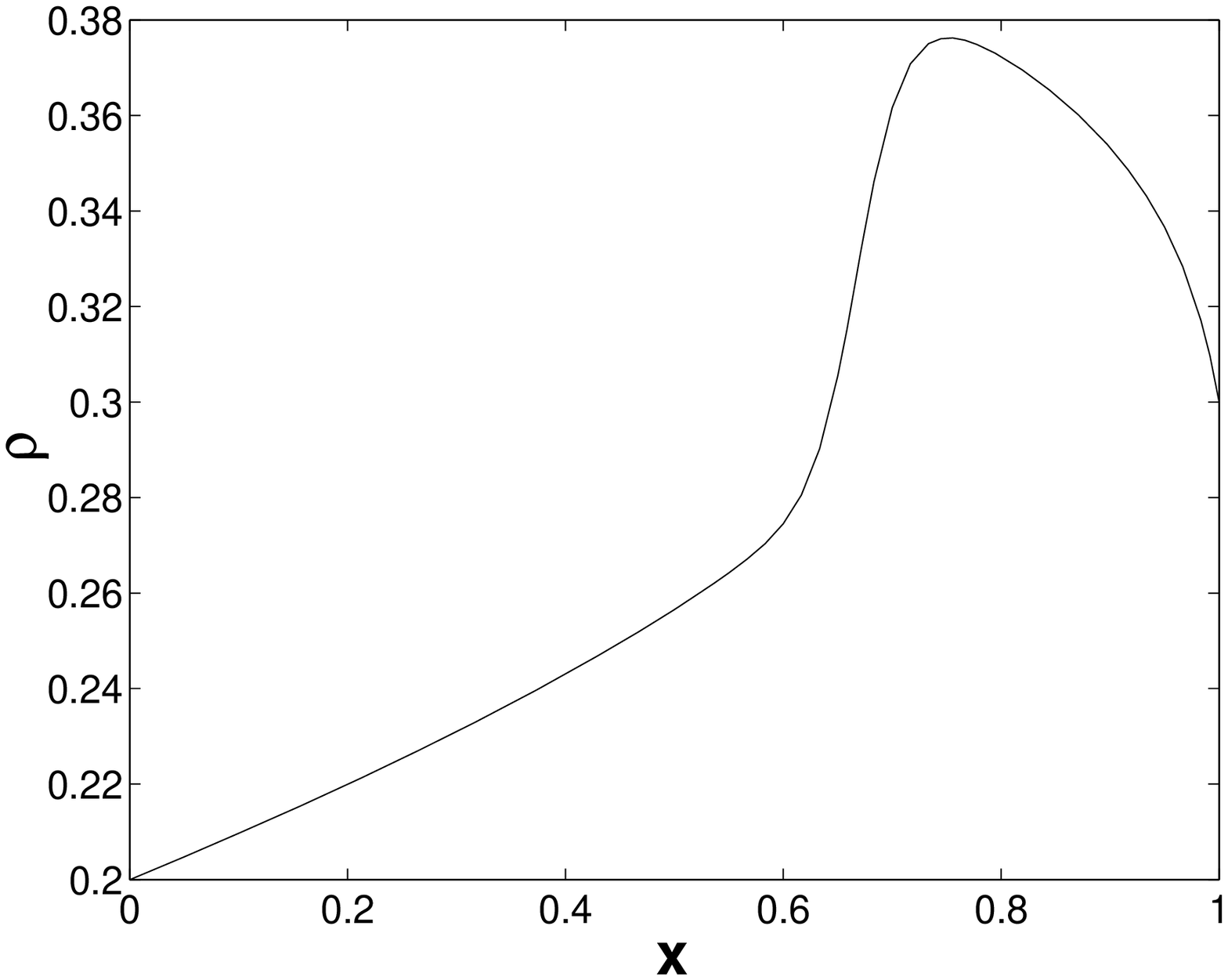}
    \caption{ Plot of the density profile for $\alpha=.20$, $\gamma=.30$,
$\Omega=.1$, $\epsilon=.009$. At these values of parameters, there is a shock 
in the density profile but it is not formed through the deconfinement 
of the boundary layer. Two branches of outer solutions are joined 
discontinuously through the shock whose height increases as $\alpha$ 
increases with fixed $\gamma$.}
\label{fig:nodeconfine}
\end{figure}
}
\begin{document}
\title{\bf{Shocks in asymmetric simple exclusion processes of 
interacting particles}}
\author{Sutapa Mukherji}
\affiliation{Department of Physics, Indian Institute of
  Technology, Kanpur 208 016, India}
\date{\today}
\begin{abstract}
In this paper, we study shocks and related transitions in  
asymmetric simple exclusion processes
of particles with nearest neighbor interactions. 
We consider two kinds of  inter-particle interactions. 
In one case, the particle-hole symmetry is broken due to the 
interaction. In the other case, particles have an 
effective repulsion due to which the particle-current-density
drops down  near the half filling. 
These interacting  particles move on an one dimensional lattice 
which is open at both the ends  with injection 
of particles at one end and withdrawal of particles at the other. 
In addition to this, there are possibilities of attachments or detachments
of particles to or from the lattice with certain rates. 
The hydrodynamic equation that involves  the exact particle
 current-density 
of the particle conserving system and additional terms 
taking  care of the attachment-detachment kinetics is studied 
 using the techniques of boundary layer analysis. 
\end{abstract}
\maketitle

\section{Introduction}
 Asymmetric simple exclusion process (ASEP) 
comprises of particles performing  biased hopping in a preferred
direction on an one-dimensional chain. This biased hopping gives rise 
to a finite particle current due to which the detailed balance 
is violated.  In its simplest  version, the particles respect 
mutual exclusion due to which a lattice site cannot be occupied by 
more than one particle.
Recently, ASEP with 
nonconservation of particles \cite{frey}
 have drawn a lot of attention because 
these systems  have better resemblance 
with the motors  participating in biological transports
inside the cell \cite{alberts} than  the particle 
number conserving systems.
The particles undergoing ASEP   are  analogous to 
the molecular motors performing directed motion on tracks laid by the 
biopolymers \cite{frey,klumpp}. 
The conservation of particle number  may be violated by  
absorption (or evaporation) of particles 
to(or from) the chain, a process, that is 
  similar to the  attachment or 
detachment of motors from the filament.

 One of the important issues involving 
these nonequilibrium processes on an 
open system  is the boundary induced phase 
transition \cite{krugprl}. Boundary induced phase 
transitions are possible for both particle conserving 
\cite{derrida} and 
nonconserving ASEP \cite{frey,evans,smbsm,smvm,smb} 
with the latter one having 
 a richer phase diagram with  new  phases.
 The other important aspect which  influences  the phase 
diagram is the interaction between the particles in addition to the 
mutual exclusion \cite{hager}.  These
interactions affect  the particle current along the lattice. Since 
the particle current  carries the boundary information 
to the bulk,  boundary induced phase transitions are likely to be 
affected by the change in the inter-particle interaction.

The  ASEP  we consider here
 is defined on an one dimensional lattice of size $l$ with $N$ 
lattice points \cite{schuetzrev}. Each site
can be empty or occupied by a particle. A particle on  site 
$i$ can hop to site $i+1$ with unit rate provided the target site is empty.
A steady flow of particles along 
this one-dimensional channel is maintained by  injection and withdrawal of 
particles at certain rates 
 at  left  and  
right ends of the lattice 
respectively. The boundaries are coupled with particle 
reservoirs with fixed particle densities. We assume that the rates are 
such that the left and the right boundaries have particle densities 
$\alpha$ and $\gamma$ respectively. 
 The particle number in the bulk is not conserved since  the 
particles are allowed to detach (or attach)  from (or to) the chain 
with rates $\omega_d$ ($\omega_a$).   
In the  steady state, depending on
the values of $\alpha$ and $\gamma$, the system is 
found to exist in different 
phases that are characterized by the particle 
density $\rho$ and the current-density 
$j(\rho)$. These phases are represented in the phase-diagram in the 
space of $\alpha$ and $\gamma$. 
 Monte Carlo simulations and mean field analysis of this model 
\cite{frey,evans} 
show low-density, high-density and maximal current 
phases in addition to a phase, known as the shock phase,  where the
low-density and high-density regions coexist. The density profile across 
the lattice has a jump discontinuity (shock)
 from a low-density region to a high-density region. It has also 
been shown that the transition to the shock phase can become critical 
under special  values of the boundary parameters \cite{frey}.
Shocks (domain walls) separating 
different steady-state density profiles 
have been observed in ASEP without the 
absorption-desorption kinetics. However, in this case, the 
position of a shock on the lattice 
fluctuates and they can, in general, move on the lattice. The dynamics 
of such shocks \cite{straley, hager}  drive the system to settle 
in one of its steady states which are known as high-density, low-density 
and maximal current phases (in case of particles having only mutual 
exclusion). Studying the
dynamics of such shocks helps in understanding 
the nature of the phase transitions in such systems. 
Unlike this particle conserving ASEP, 
here the shocks are localized \cite{frey,evans} 
and there are stationary states
 in the phase diagram where the density profiles have such shocks. 
The shock is upward in the sense that 
the low-density phase appears on the left of the shock and the high-density 
phase appears on the right of the shock.  The position of these stationary 
shocks is determined by $\alpha$, $\gamma$ and the 
absorption-desorption rates.
  A boundary layer analysis shows that the 
transition to the shock phase from a low-density phase can happen through 
a critical deconfinement of  a boundary layer from the edge of the system
\cite{smbsm}.   
In addition, it has been shown in a general way that 
this transition to the shock phase is associated 
with a dual transition \cite{smvm,smb}. The dual transition is a boundary
transition  across which 
the slope of the boundary layer  changes sign
with the shape of the bulk density profile remaining unchanged. 
These boundary and shock transition lines meet at the critical point.
The presence of  the  dual transition helps in developing  a general 
approach through which one  can classify the phase-diagrams into 
different categories. This has been done in reference \cite{smb} 
using the knowledge 
of zeros of a set of coarse grained functions present in the 
equation describing the steady state density profile.

The effect of the inter-particle interactions on the 
particle conserving ASEP 
has been found  to be quite  drastic \cite{hager}. 
Typically one considers interactions of the following nature \cite{kls}. 
The particles 
hop to the right with bulk hopping rates
\begin{eqnarray}
0 \ 1\ 0\ 0\rightarrow 0 \ 0\ 1\ 0\ \ {\rm with\ rate}\ 
1+\delta\label{partcur}\\ 
1\ 1\ 0\ 0\rightarrow 1\ 0\ 1\ 0\ \ {\rm with\ rate}\ 1+\Delta\label{rep1}\\
0\ 1\ 0\ 1\rightarrow 0\ 0\ 1\ 1\ \ {\rm with\ rate}\ 1-\Delta\label{rep2}\\
1\ 1\ 0\ 1\rightarrow 1\ 0\ 1\ 1\ \ {\rm with\ rate}\ 1-\delta.\label{vaccur}
\end{eqnarray}
Here, $1$ and $0$ imply an occupied and an unoccupied site respectively.
$\Delta=\delta=0$, is the  case of ASEP with only mutual exclusion. 
In this case, the   current-density  
 $j(\rho)=\rho(1-\rho)$ respects particle-hole symmetry
and has a maximum at  $\rho=1/2$. 
An exact solution of the particle conserving 
problem \cite{derrida} 
reveals that there are primarily 
three phases known as the low-density, high-density and the maximal 
current phases. 
The  case of  $\Delta\rightarrow 1$ 
implies a repulsion among the particles due to 
which the current versus density 
plot is  expected to show a minimum near the half-filling. 
In the range $0\le \Delta\le 1$, as $\Delta$ increases,
 the current-density plot 
 develops  double  maxima from a single maximum picture.
With a minimum in the current versus
 density plot, the phase diagram contains 
a minimum-current phase  \cite{hager}
 in addition to the low-density, high-density  and 
the maximal current phases. At this minimum current phase, the bulk 
density has a value that corresponds to the minimum value of the 
current.   The maximal-current phase,
 in general, is also expected to be of two different types  in which 
 the bulk density corresponds to two different maxima of the current.  
  Equations \ref{partcur} and \ref{vaccur} involving $\delta$
 imply that the particle current is the same 
 as the vacancy current for $\delta=0$.  
A finite $\delta$ breaks this equality since $\delta> 0$ 
and $\delta< 0$ imply larger particle or vacancy 
currents respectively. This 
particle-hole asymmetry introduces an asymmetry in the current
versus density plot.

Since the particle nonconservation and the inter-particle interaction both 
individually leave significant impact on ASEP, their combined effect is 
expected to be interesting. 
Although the phase diagram of the particle nonconserving version of the 
interacting  ASEP
has not   been probed,   numerical 
simulations exhibit some new  features such as downward shock (see figure 
 \ref{fig:downshock}) and  
 double shocks \cite{popkov} 
in the density profile. 
\picturedownshock
In case of downward shock, 
the shock is a discontinuity in the density 
profile separating a  high-density part 
on the left  and a low-density 
part on the right. The density profile with a double shock consists of 
two successive upward jumps separating a low-density part on the 
left and a high-density part on the right.    
The particular feature that interests us here is the downward shock in the 
density profile. 

 In this paper, we look into the details of the  shocks for two limiting 
cases 
(i) $\Delta=0$ and $\delta=1$ and (ii) $\Delta=1$ and $\delta=0$ 
using a boundary layer analysis. These problems serve as the limiting cases 
of the most general interacting case which is technically more difficult to 
handle. The first case is  useful for the  verification  of our earlier 
predictions 
related to shockening transition, dual boundary transition and criticality. 
We show that although the shape of the density profile 
is different in this case, different phase transitions and the 
phase diagram are qualitatively 
same as those of  only mutual exclusion case
with $\omega_a \neq \omega_d$. 
 Further, the asymmetry in the current versus 
density plot does not alter the 
exponents which remain the   same as those
of the pure mutual exclusion case with $\omega_a \neq \omega_d$. 
In the second case,     
our focus is particularly on the shape 
of the  density profile that has a 
downward shock, location of the shock and   specific 
conditions on the boundary parameters $\alpha$ and $\gamma$ for which
such shock can be observed. The aim is also to show how the earlier analysis 
for the upward shock can be appropriately modified to understand the
details of the downward shock. We show that the earlier predictions
 \cite{popkov} 
about the ranges in which the values of the  densities at the upper and lower 
end of the shock must lie, are satisfied
 naturally  in the boundary layer analysis. 
 Although the boundary layer analysis presented here can be used 
 to obtain other features such as  phases with  upward shocks,  
entire phase diagram with the details of the phase boundaries, 
we plan to present those with supportive  numerical 
analysis of the system in a later publication.

The paper is organized as follows. In the next section, we discuss 
some of the 
details of the model. In  section III, we consider the first case,
namely that of $\Delta=0$ and $\delta=1$. Some results related to the 
density profiles and phase diagrams for this case are presented in the 
Appendix. In section IV, we consider the case of $\Delta=1$ and $\delta=0$.
We conclude the paper with a discussion in section V.

\section {Model}

In general, for a
 particle conserving system, one can write down a discrete
continuity equation 
\begin{eqnarray}
\frac{d\rho_k}{dt}=j_{k-1}-j_k,\label{discrete}
\end{eqnarray}
where $j_k$ is the average current across the bond between the $k$th and 
$k-1$th sites and $\rho_k$ is the average particle occupancy at the $k$th 
site. The average current takes care of the particle hopping rule that 
is problem specific. In the presence of  particle 
absorption-desorption kinetics, we may supplement this equation with other 
terms responsible for particle absorption and desorption. One can obtain a
simple, solvable continuum model from the resulting equation 
by making a mean field approximation that neglects  particle correlations.
 In  case of only mutual exclusion, 
the  current-density obtained from the 
mean-field  approach is the exact one and as a consequence of this there is 
a good  agreement in results obtained from the  
mean-field analysis  
and the Monte-Carlo simulation. The  mean-field approach, however,  
does not produce  correct results in the 
 presence of additional interactions
considered here. 
This is because the mean-field approach fails to produce even the basic 
qualitative features of the current-density.  
  
In view of these issues, we follow the  approach 
of hydrodynamic equation \cite{popkov}. 
In this approach,  we start with the 
thermodynamic limit, $N\rightarrow \infty$
with the lattice spacing $a=l/N\rightarrow 0$.
For the entire analysis in the following, we choose $l=1$.
In the continuum limit, $k\rightarrow x=ka$ and 
$t_{\rm lattice}\rightarrow t=t_{\rm lattice} a$,
the average particle density $\rho(x,t)$ satisfies the 
continuous version of the continuity equation.
 With the particle 
nonconserving terms as additional source and sink terms, we have the 
continuous  equation 
\begin{eqnarray}
\frac{\partial{\rho}}{{\partial t}}+\frac{\partial}{\partial x} j(\rho)
=\Omega_a(1-\rho(x,t))-\Omega_d \rho(x,t).\label{hydro}
\end{eqnarray} Here  
$\Omega_a=\omega_a N$ and $\Omega_d=\omega_d N$. We use the exact
 stationary  current in the hydrodynamic equation 
with the assumption that the bulk 
has sufficient time to relax between the particle absorption and 
desorption events. This is possible when the 
absorption and desorption happen at a very low rate. 
 
The current-density for the interacting system has been derived 
in the past \cite{hager}
 by using the fact that the stationary distribution is 
given by the 
equilibrium distribution of an one dimensional Ising model. 
The average 
current across the bond between $k$ and $k+1$th sites is given by 
\begin{eqnarray}
&& j_k=(1+\delta)\langle 0100\rangle+(1+\Delta)\langle1100\rangle+
\nonumber\\
&& (1-\Delta)\langle 0101\rangle+(1-\delta)\langle 1101\rangle,
\end{eqnarray}  
where the four sites in the equation 
are the $k-1,\ k,\ k+1,\ k+2$th sites. The four-point 
averaging needs to be done with respect to the stationary measure. 
Since the details of the calculations are available in \cite{hager}, 
we quote the final result for the current-density. 
In the thermodynamic limit, $N\rightarrow \infty$,  the 
current-density is
\begin{eqnarray}
j=\frac{\lambda[1+\delta(1-2\rho)]-\Delta \sqrt{4\rho(1-\rho)}}{\lambda^3},
\end{eqnarray}  
where 
\begin{eqnarray}
\lambda=\frac{1}{\sqrt{4\rho(1-\rho)}}+(\frac{1}{4\rho(1-\rho)}-1+
\frac{1-\Delta}{1+\Delta})^{1/2}.
\end{eqnarray}
For the  mutual exclusion case, $\Delta=\delta=0$, one obtains 
the exact current-density  
$j=\rho(1-\rho)$. This is also the current-density one finds through a 
mean-field approach 
starting with $j_k=\langle 10\rangle$. Mean-field amounts to 
ignoring the correlation and assuming 
$j_k=\langle 1\rangle \langle 0\rangle$. 
For $\delta=1$ and $\Delta=0$, 
the current-density is $j=2 \rho(1-\rho)^2$. The  current-density 
vanishes in the completely unoccupied, $\rho\rightarrow 0$, and 
completely occupied, $\rho\rightarrow 1$, limits with a maximum at 
$\rho=1/3$. The lack 
of the particle-hole symmetry is reflected in the asymmetric 
nature of the current-density around 
$\rho=1/2$. 
For $\Delta=1$ and $\delta=0$, the current-density is 
\begin{eqnarray}
j=y(1-y)/(1+y),
\end{eqnarray}
where 
\begin{eqnarray}
y=\sqrt{1-4 \rho(1-\rho)}.
\end{eqnarray}
The current-density is  symmetric about $\rho=1/2$, and has 
 two maxima at $\rho=.707$ and $\rho=.293$ 
(see figure \ref{fig:callcurrentplot}).
\currentdensity

Equation (\ref{hydro}) is  subjected to the 
boundary conditions $\rho(x=0)=\alpha$ and $\rho(x=1)=\gamma$.
Although the first order equation obtained from equation (\ref{hydro}) 
under the steady-state condition, $\partial\rho(x,t)/\partial t=0$, 
describes  the shape 
of the steady-state density profile as a function of $x$, this equation, 
in general, does not have  a smooth solution satisfying two 
boundary conditions. 
 This problem can be overcome 
by adding a second order 
term  to this equation. This second order term is  
introduced with  a prefactor which becomes vanishingly small in the 
thermodynamic limit. It can be  shown that such an infinitesimal 
second order term is generated anyway as one takes 
 the continuum limit of (\ref{discrete}) and retains terms up to 
$O(a)$.
 The  second order term ensures a smooth 
solution for the hydrodynamic equation through the formation of 
shocks or boundary layers over a region of width of the order 
of the lattice spacing, $1/N$. 
In the steady-state,  
the final form of the equation describing the shape of the density 
profile is, therefore,
\begin{eqnarray}
\frac{\epsilon}{2}\frac{d^2\rho}{dx^2}-\frac{dj}{dx}+
\Omega_a (1-\rho(x))-\Omega_d \rho(x)=0,\label{stationary}
\end{eqnarray}
where $\epsilon$ is a factor of the $O(1/N)$ and 
$\epsilon\rightarrow 0$ in the 
thermodynamic limit, $N\rightarrow \infty$.  
In the presence of the 
second order term, one can define an  effective current as 
\begin{eqnarray} 
J=-\frac{\epsilon}{2} \frac{d\rho}{d x}+j.
\end{eqnarray} We discuss later the important role $J$ plays 
in predicting the nature of the density profile.

A knowledge of certain special densities obtained from 
equation (12) can be useful for us. 
In the presence of only  absorption-desorption kinetics, 
with $K=\Omega_a/\Omega_d$, the 
system settles into a stationary  density, 
$\rho=\rho_L=K/(K+1)$  obtained from 
the zero of the particle non-conserving terms. It is possible for 
the system to settle into this constant equilibrium density profile
in the presence of hopping dynamics also 
 provided the equation of motion allows such a density profile. 
This is what happens in case of ASEP with only mutual exclusion. 
In this case, the 
current-density $j(\rho)=\rho(1-\rho)$  has a 
maximum ($j'(\rho)\mid_{\rho_m}=0$)at $\rho=1/2$
and with $K=1$, one has 
$\rho_L=\rho_m$. This is a very special situation which 
gives rise to a  phase
in which the  density profile can maintain the maximum current 
in the system. Thus for $K=1$, in addition to the low-density, 
high-density and  shock 
phases one also has a maximal current phase and 
coexistence of the maximal current phase with other phases.   
Different phases and  different 
exponents associated with the phase transitions in the 
purely mutual exclusion case   
indicate the existence of a different universality class for $K=1$.  
A recent paper \cite{smb}  shows that instead of the value of $K$, it 
is the relative magnitude of 
$\rho_L$ and $\rho_m$ that determines the phase diagram of the system.
 While $\rho_L=\rho_m$ ( for purely 
mutual exclusion case, this condition is fulfilled if $K=1$) 
forms a special case where the 
maximal current phase appears, the phase diagrams for 
$\rho_L<\rho_m$ and $\rho_L>\rho_m$ are 
related through the particle-hole 
symmetry. In our entire analysis, we consider $K=1$ 
($\Omega_a=\Omega_d=\Omega$). This, however, 
does not restrict us to any special universality class as we maintain 
$\rho_L\neq \rho_m$.

\section{Basic principles of Boundary layer analysis}
We mention here a few basic principles that have been employed 
in the following boundary layer analysis. We start from the basic 
hydrodynamic equation   (\ref{stationary}). 
Since the contribution of the second order term in 
this equation  is very small in the thermodynamic limit,
it is expected that the major part 
of the density profile is described by the solution  
of the first order equation obtained from  equation 
(\ref{stationary}) by ignoring  the second  order 
term. This solution, known as the outer solution, 
cannot satisfy both the boundary conditions in general. 
In order to satisfy 
the boundary conditions appropriately, there appear  special narrow 
regions (of width of $O(\epsilon)$) 
such as boundary layers or shock. In order to see these regions, it is 
necessary to rescale the position variable of the differential 
equation appropriately. These regions are described 
by inner solutions which are solutions of  the rescaled 
equation.  Upon rescaling equation (\ref{stationary}), the 
nonconservative terms  acquire a prefactor of $O(\epsilon)$, due to which 
these terms are neglected.  
The inner solution is, therefore, determined by the 
first two terms of equation (\ref{stationary}). 
  Different unknown constants, 
present in the solutions in different regions,  are determined   
from the boundary conditions or by the conditions required for 
 the smooth 
joining of the outer and inner solutions. This is the general scheme 
\cite{cole}
through which one can obtain an uniform approximation of the 
solution of equation (\ref{stationary}) in the thermodynamic limit. 

Based on these  principles, a few predictions about the presence 
or absence of the shock in the density profile can be made.
The knowledge about the  shape of the density profile with shock or the 
location of the shock,
 however, requires explicit calculations.  
Since the particle nonconserving terms of equation (\ref{stationary}) are 
 not important for the boundary layer
or the shock region, we expect the total current $J$ to remain constant in 
this region. For a shock separating two densities $\rho_{\rm lo}$
 and $\rho_{\rm ro}$, 
we have the total current $J=j(\rho_{\rm lo})$ 
and $J=j(\rho_{\rm ro})$ at the two edges 
of the shock and   the  constancy of the 
total current $J$ across the shock, demands 
$j(\rho_{\rm lo})=j(\rho_{\rm ro})$. 
This condition is implemented explicitly 
in the boundary layer analysis in 
finding out the inner solutions describing the shocks or 
boundary layers.  
As a consequence of this constancy condition, 
the path representing the shock on the 
current-density plot is horizontal.  
 For an upward shock in the 
density profile ($d\rho/dx>0$), one requires $J<j(\rho(x))$. The horizontal 
path representing the 
upward shock on the current-density plot 
has to lie below the curve
  $j(\rho(x))$. Therefore,  we require at least one maximum in the 
current and density relation  to fulfill these conditions 
 for  an upward shock. 
Similarly, we expect $j(\rho(x))<J$ in case of a downward shock 
($d\rho/dx<0$). This requires a concave region in the current versus 
density plot. In our case, due to the particle-hole symmetry, the center 
of the downward shock lies at $\rho=1/2$. As a result, $\rho_{\rm lo}$ and 
$\rho_{\rm ro}$  are bounded as $.5<\rho_{\rm lo}<.707$ and 
$.293<\rho_{\rm ro}<.5$. 
In the boundary layer analysis, these conditions appear 
naturally as requirements for the saturation of the shock solution 
(inner solution) to the bulk on both sides of the shock.

\section{Boundary layer analysis for 
$\delta=1$ and $\Delta=0$}

We start our analysis by finding out the outer solutions.
The outer solution 
is the solution of equation 
\begin{eqnarray}
-2(1-4 \rho+3 \rho^2) \frac{d\rho}{dx}+\Omega(1-2\rho)=0,
\end{eqnarray}
obtained from equation  (\ref{stationary}) 
by ignoring the second order term.
The solution is given in terms of the transcendental equation 
\begin{eqnarray}
g(\rho_{\rm out})=\Omega x+c,
\end{eqnarray} 
where 
\begin{eqnarray}
g(\rho)=
-\frac{1}{2} (3\rho^2-5 \rho)+\frac{1}{4} \log[2\rho-1]
\end{eqnarray} and 
 $c$ is a constant that can be  found out from the  boundary 
condition that the outer solution satisfies. 
As in the pure mutual exclusion case, there can be a phase 
where the outer solution satisfies the left boundary condition 
$\rho(x=0)=\alpha$. 
We consider this situation in the following. In this case, the outer 
solution is given by  
\begin{eqnarray}
g(\rho)=\Omega x+g(\alpha).\label{outer2}
\end{eqnarray}

The inner solution describing the boundary layers or shocks can be 
found from equation (\ref{stationary}), by expressing it in terms of 
$\tilde x=(x-x_0)/\epsilon$, where $x_0$ represents the location of the 
inner solution. The nonconservative terms in the rescaled 
equation are negligible in the $\epsilon\rightarrow 0$ limit. The 
inner solution is, therefore, the solution of the equation 
\begin{eqnarray}
\frac{1}{2} \frac{d\rho}{d\tilde x}=2(\rho-2 \rho^2+\rho^3)+C,\label{innereqn1}
\end{eqnarray}
where $C$ is a constant. The saturation of the 
inner solution to  $\rho_{\rm o}=\rho_{\rm out}(x\rightarrow 1)$ as 
$\tilde x\rightarrow -\infty$ is ensured by choosing 
\begin{eqnarray}
C=-2(\rho_{\rm o}-2 \rho_{\rm o}^2+\rho_{\rm o}^3).
\end{eqnarray}
Equation (\ref{innereqn1}) can be rewritten as  
\begin{eqnarray}
\frac{d\rho}{d\tilde x}=4(\rho-\rho_{\rm o})(\rho-\rho_1)(\rho-\rho_2),
\label{innereqn2}
\end{eqnarray}
where 
\begin{eqnarray}
\rho_{1,2}=\frac{1}{2}
[2-\rho_{\rm o}\pm\{4\rho_{\rm o}-3\rho_{\rm o}^2\}^{1/2}].\label{satval} 
\end{eqnarray}
The general solution of   equation (\ref{innereqn2}) is 
\begin{eqnarray}
\log[\frac{(\rho-\rho_{\rm o})^{(\rho_1-\rho_2)}
(\rho-\rho_1)^{(\rho_2-\rho_{\rm o})}}{(\rho-\rho_2)^{(\rho_1-\rho_{\rm o})}}]
=\nonumber\\
4{(\rho_{\rm o}-\rho_1)(\rho_{\rm o}-\rho_2)(\rho_1-\rho_2)}
(\tilde x+\xi),\label{gensolnin}
\end{eqnarray}
where $\xi$ is a constant. 
There are  possibilities of saturation of the inner solution to
densities $\rho_1(\rho_{\rm o})$ or $\rho_2(\rho_{\rm o})$ 
as $\tilde x\rightarrow \infty$. These two densities are functions of 
$\alpha$ and $\Omega$ through $\rho_{\rm o}$. 
 The possibility of saturation to $\rho_1$ 
can be ruled out since for all values of $\rho_{\rm o}$ in the 
range $\{0,1\}$, $\rho_1>1$. 
The inner solution can, however, saturate to $\rho_2$ which remains 
in the realistic range ($<1$) in the entire range of $\rho_{\rm o}$.
In case of saturation, the approach to $\rho_2$ is given by 
\begin{eqnarray}
\rho\sim \rho_2+&& 
(\rho_2-\rho_{\rm o})^{(\rho_1-\rho_2)/(\rho_1-\rho_{\rm o})}
(\rho_2-\rho_1)^{(\rho_2-\rho_{\rm o})/(\rho_1-\rho_{\rm o})}\nonumber\\
 && \exp[-4
(\rho_2-\rho_{\rm o})(\rho_1-\rho_2)(\tilde x+\xi)].\label{in-sat}
\end{eqnarray}
Two length scales  appear  in the inner solution. 
The length scale $\xi$ describes the center of the inner solution. 
Using the boundary condition, $\rho(\tilde x=0)=\gamma$, 
that equation (\ref{gensolnin}) must satisfy, we have 
\begin{eqnarray}
\xi=&& \frac{1}{4(\rho_{\rm o}-\rho_1)(\rho_{\rm o}-\rho_2)
(\rho_1-\rho_2)}\times\nonumber\\ && \log[\frac{
(\gamma-\rho_{\rm o})^{\rho_1-\rho_2} (\gamma-\rho_1)^{\rho_2-\rho_{\rm o}}}
{(\gamma-\rho_2)^{\rho_1-\rho_{\rm o}}}].
\end{eqnarray}
The other length scale, which we denote as  $w$,
 describes the approach of the 
inner solution to the saturation value. This length scale is 
\begin{eqnarray}
w=1/[4(\rho_1-\rho_2)(\rho_2-\rho_{\rm o})].
\end{eqnarray}
This case of  outer  and inner 
solutions  satisfying 
the boundary condition at $x=0$ and  $x=1$ respectively is  
possible for low values of $\alpha$. 
This phase, 
with $\alpha$ dominated bulk density for low values of $\alpha$ is the
 low-density phase.

 The boundary layer at $x=1$ is unable to satisfy the 
right boundary condition if $\gamma>\rho_2$  and  as a consequence of this, 
the boundary layer 
deconfines from the boundary at $x=1$ as $\gamma>\rho_2$. To satisfy the 
right boundary condition, an outer solution joined smoothly to the boundary 
layer at the left and satisfying the right boundary condition at  its right 
appears. The deconfinement of the boundary layer from the boundary 
 as $\gamma$ exceeds $\rho_2$ is the shockening transition that has been 
 seen earlier in the pure mutual exclusion case.
The phase boundary between the low-density and the shock phase
 can be 
determined from the condition 
\begin{eqnarray}
\rho_2(\alpha)=\gamma.
\end{eqnarray} 
As the shock phase boundary is approached the length scale $\xi$ diverges
logarithmically as 
\begin{eqnarray}
\xi\sim \log(\gamma-\rho_2).
\end{eqnarray}

Equation (\ref{innereqn2}) also exhibits a boundary transition 
 that has been  first identified as a dual transition 
in the pure mutual exclusion case \cite{smvm} 
and is expected to be present in general 
whenever there is a bulk shockening transition through a deconfinement of 
a boundary layer.  This boundary transitions happens in the 
low-density phase and across this boundary transition, the slope 
of the boundary layer (inner solution) changes sign. It can be seen from  
equation (\ref{innereqn2}), that the boundary layer has   a 
positive slope if $\rho_{\rm o}<\gamma<\rho_2$ and has a negative slope  
if $\gamma<\rho_{\rm o}$. Thus, this change in the  slope of the 
boundary layer  happens across the  boundary transition line 
\begin{eqnarray}
\rho_{\rm o}(\alpha)=\gamma.
\end{eqnarray}
We call the low-density phase with boundary layer having positive 
slope as low-density(1). The other part of the low-density phase is 
called as low-density(2).  
 The boundary transition 
 supports the  duality theorem of reference \cite{smvm} 
that for every $\alpha$ if there is a bulk phase transition at 
$\gamma=\rho_{\rm 2}$, there is a boundary transition at 
$\gamma=\rho_{\rm o}(\alpha)$.
As the boundary transition line is approached from either side, the length
scale $\xi$ diverges logarithmically as
\begin{eqnarray}
\xi\sim \log(\rho_{\rm o}-\gamma).
\end{eqnarray}
The bulk shockening transition line and the boundary transition line 
meet at a point where 
\begin{eqnarray}
\rho_{\rm o}=\rho_2=\gamma.\label{condcrit}
\end{eqnarray} This is the critical 
point at which the length scale $w$ diverges as 
\begin{eqnarray}
w\sim \frac{1}{(\rho_2-\rho_{\rm o})}.
\end{eqnarray}
Equation (\ref{condcrit}), along with equation (\ref{satval}) 
leads to the critical value for $\gamma_c=1/3$. The corresponding 
critical value of $\alpha$ can be determined by implementing 
the condition 
$\rho_{\rm o}(\alpha_c)=1/3$ 
in equation (\ref{outer2}). This leads to the following equation
\begin{eqnarray}
g(1/3)=\Omega+g(\alpha_c),
\end{eqnarray}
from which the $\Omega$ dependent $\alpha_c$ can be determined.

The analysis presented here  agrees with  the predictions of 
our earlier work. Although numerical values of $\alpha$ and $\gamma$ for 
the critical point or for other transitions are now shifted, the 
qualitative features of the phase diagrams and phase boundaries 
remain same as that of 
the pure mutual exclusion  case 
with $\Omega_a\neq \Omega_d$. We have shown here that 
the nature of divergences
of different length scales associated with the shock transition and the 
boundary transition are same as those in the case of ASEP 
with only mutual exclusion \cite{smbsm,smvm}. 
Since the calculation of other exponents 
are similar to those of ASEP with only mutual exclusion,
we refer the 
reader to references \cite{smbsm,smvm} for this purpose.

\section{Boundary layer analysis for  
$\delta=0$ and $\Delta=1$}
For $\delta=0$ and $\Delta=1$,
we look for those solutions of equation (\ref{stationary}) 
which support a downward shock in the density profile. 
In this system, the    downward shock is seen due to the presence 
of a concave region in  
the current versus density plot. Due to the particle-hole symmetry, 
we expect the downward shock to be 
centered around $\rho=1/2$. Further, there is a discontinuity in $dj/d\rho$ 
at $\rho=1/2$, because of which 
we need to distinguish two regions, $\rho>1/2$ and $\rho<1/2$.
As a consequence, the boundary layer analysis for the density profile
differs significantly from that of the asymmetric case considered in  
the first part of this paper and also from 
our previous studies in \cite{smbsm,smvm}

  The bulk solutions on the left 
and right of the downward shock are described by the left and right outer 
solutions obtained below.    
The outer solutions  
can be obtained by solving equation (\ref{stationary})
 with its first term ignored. The left outer solution, for which  $\rho>1/2$,
 is the solution of 
\begin{eqnarray}
-\frac{2(1-2 y-y^2)}{(1+y)^2}
\frac{d\rho}{dx}+\Omega(1-2\rho)=0,
\end{eqnarray} 
with  $y=2 \rho-1$. 
The solution of the equation is 
\begin{eqnarray}
g_l(\rho_{\rm lout})=\Omega x+c_1,\label{outerdown1}
\end{eqnarray}   
where
\begin{eqnarray}
g_l(\rho)=-\frac{1}{\rho}+2\log\rho-\log(2\rho-1).
\end{eqnarray}
 $c_1$ can be fixed from the boundary condition, $\rho(x=0)=\alpha$, 
since the left outer solution satisfies the left boundary condition. 

The equation for the right outer solution can be found out 
from (\ref{stationary}) in a similar way with the substitution  
$y=1-2\rho$. 
The right outer solution on the right of the downward shock 
is 
\begin{eqnarray}
g_r(\rho_{\rm rout})=\Omega x+c_2,\label{outerdown2}
\end{eqnarray}
where 
\begin{eqnarray}
g_r(\rho)=-\frac{1}{\rho-1}-2 \log(\rho-1)+\log(2\rho-1).
\end{eqnarray}
As before, $c_2$ can be determined from the condition $\rho(x=1)=\gamma$. 
The plot of these two solutions for given $\alpha$ and $\gamma$ 
is shown below.
\outersoln

In the following, we find out the inner solutions that describe 
the downward shock.  From the symmetry 
of the current-density plot about  $\rho=1/2$, 
we expect  the  center of the inner solution to correspond to  
$\rho=1/2$. The inner solution that approaches the left outer 
solution can be obtained from  equation (\ref{stationary}) with $y=2\rho-1$, 
after expressing it in terms of $\tilde x=(x-x_s)/\epsilon$, where $x_s$ 
represents the center of the inner solution. With the rescaling 
of $x$, the $\Omega$ dependent term in equation (\ref{stationary}) drops 
out for having negligible contribution. 
The  equation that determines  the left inner solution is  
\begin{eqnarray}
\frac{d\rho}{d\tilde x}=-\frac{2+4\rho^2}{\rho}+d_1.\label{inner1}
\end{eqnarray}
We expect the inner solution to saturate to  the left outer solution 
$\rho_{\rm lo}=\rho_{\rm lout}(x\rightarrow x_s-)$ as 
$\tilde x\rightarrow -\infty$.  Therefore, we choose 
\begin{eqnarray}
d_1=\frac{2+4\rho_{\rm lo}^2}{\rho_{\rm lo}}.
\end{eqnarray}
The solution of (\ref{inner1}) is given by the transcendental equation
\begin{eqnarray}
-\frac{\rho_{\rm lo}^2}{2}\log(\rho-\rho_{\rm lo})+
\frac{1}{4}\log(2\rho\rho_{\rm lo}-1)=(\tilde x+c)(2\rho_{\rm lo}^2-1).
\label{leftin}
\end{eqnarray}
The constant $c$ can be found out by 
demanding $\rho(\tilde x=0)=1/2$. The inner solution should approach 
$\rho_{\rm lo}$ as $\tilde x\rightarrow -\infty$.
This is possible if  $2\rho_{\rm lo}^2-1<0$. Thus on the high-density 
side, the shock should saturate to a density $\rho_{\rm lo}$ 
which is bounded as  
\begin{eqnarray}
.707>\rho_{\rm lo}>.5. \label{bound1}
\end{eqnarray} 
The right inner solution that approaches the right outer solution 
$\rho_{\rm ro}=\rho_{\rm rout}(x\rightarrow x_s+)$ as 
$\tilde x\rightarrow \infty$ can be found out in a similar way. 
Substituting  $y=1-2\rho$ 
in equation  (\ref{stationary}), we find the right inner solution 
\begin{eqnarray}
\frac{p_{\rm ro}^2}{2} \log(p-p_{\rm ro})-\frac{1}{4}\log(2 p p_{\rm ro}-1)
=(2 p_{\rm ro}^2-1)(\tilde x+d),\label{rightin}
\end{eqnarray}
where $p=1-\rho$. 
The constant $d$ can be found out by demanding that $p=1/2$ at $\tilde x=0$.
This solution also saturates to $\rho_{\rm ro}$ as $\tilde x\rightarrow 
\infty$,
with the condition that $2 p_{\rm ro}^2-1<0$. Thus on the low-density 
side, the shock saturates to a value bounded as 
\begin{eqnarray}.29<\rho_{\rm ro}<.5.\label{bound2}
\end{eqnarray} 
Conditions in (\ref{bound1}) and (\ref{bound2}) are same as those 
predicted for a downward shock for the 
particle conserving interacting systems. 
The boundary layer analysis presented here 
 shows that these conditions are necessary 
for the saturation of the inner solutions. 

In the fully symmetric situation ($\delta=0$), the inner solution is 
centered at $\rho=1/2$. This  imposes another constraint 
\begin{eqnarray}
\rho_{\rm lo}=1-\rho_{\rm ro}.\label{slope}
\end{eqnarray} 
This condition also guarantees
the matching of the slopes of the left and right inner solutions 
at $x=x_s$. 
Equation (\ref{slope})
allows us to find out the shock positions and the
 dependence  of $\rho_{\rm lo}$ and $\rho_{\rm ro}$ on 
$\alpha$ and $\gamma$. 
If the shock is formed at $x=x_s$, we have 
\begin{eqnarray}
g_{\rm l}(\rho_{\rm lo})=\Omega x_s+g_{\rm l}(\alpha)\label{rholo}\\
g_{\rm r}(\rho_{\rm ro})=\Omega x_s+g_{\rm r}(\gamma)-\Omega.\label{rhoro}
\end{eqnarray}
Using equations (\ref{rholo}), (\ref{rhoro}) and 
(\ref{slope}), we have the final equation,
\begin{eqnarray}
g_{\rm l}(\rho_{\rm lo})=g_{\rm r}(1-\rho_{\rm lo})+\Omega+
[g_{\rm l}(\alpha)-g_{\rm r}(\gamma)], \label{condition}
\end{eqnarray}
that determines $\rho_{\rm lo}$.
Equation (\ref{condition}) and (\ref{bound1}) together set the 
condition for the formation of the shock. 
Knowing the value of $\rho_{\rm lo}$ from equation (\ref{condition}) 
for given values of $\alpha$, $\beta$ and $\Omega$, 
one can find out the height 
of the shock 
 $H_{\rm shock}=\rho_{\rm lo}-\rho_{\rm ro}=2\rho_{\rm lo}-1$.

\section{Summary}
In this paper, we present a boundary layer analysis for studying the shocks 
and associated transitions for asymmetric simple exclusion processes 
of interacting particles. The particles that hop in a particular direction 
 on a finite
 one-dimensional lattice interact mutually. These interactions are present 
in addition to the usual mutual exclusion among the particles. We consider 
two different kinds of interactions. In one case, particles have mutual 
repulsion.  In the other 
case, particle-hole symmetry is broken by the interaction. We consider 
these two cases separately since in that case simple analytical solutions 
for the density profile with shock can be obtained. 
 In addition to these interactions, 
there are possibilities of attachment(detachment) of particles to(from) 
the lattice. A steady flow of particles on the lattice is maintained 
through the injection of particles at one end and withdrawal of particles 
at the other end at certain 
 rates. We assume that the rates are such that 
the particle densities at the two ends remain $\alpha$ and $\gamma$. 
 Depending on the values of these 
boundary densities $\alpha$ and $\gamma$, the system, in the steady state, 
 can be in different phases. These 
phases are characterized by the shapes of the particle 
density profiles across 
the lattice and the particle-current-densities. 
The hydrodynamic equation that 
involves the exact current-density of the particle conserving system 
is supplemented with additional particle absorption and desorption terms.
In our entire analysis, we choose equal particle absorption and 
desorption rates.
The resulting  equation is, then, studied using the techniques of 
boundary layer analysis to obtain the 
steady-state density profiles.

In the case where the interaction breaks the particle-hole symmetry, the 
phase diagram and the phase transitions are qualitatively similar to that 
of the purely mutual exclusion case with unequal particle 
absorption-desorption rates. In our interacting 
 system, the particle density profile 
shows a jump discontinuity or shock 
from a low value to a high value over an extended region on 
the $\alpha$-$\gamma$ space.
 As in the case of ASEP with only mutual exclusion, the
transition to the shock phase happens through the deconfinement of the 
boundary layer. Associated to this transition to the 
shock phase, there is also 
a dual boundary transition across which the 
slope of the boundary layer changes
sign with the bulk density profile remaining the same. The exponents 
characterizing various phase transitions are same as those of the
 ASEP with only mutual exclusion of particles. 

In case of repulsion among the particles, it is known that there exists 
a downward shock for certain values of the boundary densities. 
We obtain the analytical form of the density profile that has a 
downward shock. The solutions 
describing the shock region in the density profile saturate to the bulk part 
of the density profile exponentially. This exponential approach is possible 
if the bulk density at the edges of the 
shock remain within certain range.
We further obtain conditions on the boundary parameters for having such 
a downward shock, and
 also the location and the height of the shock. 
This work  opens up the scope of analyzing more 
general  interacting problems. Even in the absence of any compact 
 closed form analytical solutions for the density profile,
 the basic principles 
used in this problem should be applicable in  more general cases.

\appendix
\section{density profiles at various phases and phase diagram for 
$\delta=1,\ \Delta=0$}
In this appendix, we present a few representative plots of the density 
profile at different phases. 
The density profile in the 
low-density phase (low-density(1)) has the shape 
as given in figure \ref{fig:justlowdensity}.
\lowdensity
\shocktrans
Figure \ref{fig:justdeconfine} shows the deconfined boundary layer 
as $\gamma$ is increased from the shock transition value  $\gamma=.43..$ 
for $\alpha=.15$.

The deconfinement of the boundary layer to form the shock happens only 
for $\gamma>\gamma_c$. For $\gamma<\gamma_c$, 
as $\rho_{\rm o}$ becomes equal to  
$1/3$ with the increase in $\alpha$, the shock starts forming with 
an outer solution appearing on the right of the shock and satisfying an 
effective right boundary condition  $\rho(x=1)=\gamma_c$. The true 
boundary condition at $x=1$ is satisfied by a decaying boundary layer
similar to the one present in the low-density phase.
Since for all $\gamma<\gamma_c$, 
the effective boundary condition on the right 
for shock formation remains same, the low-density-shock phase boundary 
is vertical at $\alpha=\alpha_c$ for $\gamma<1/3$. 
The shock height increases from zero continuously as one enters the shock 
phase. The density profile typically appears as in figure 
\ref{fig:nodeconfine}.
\differshock

The shock, after being formed at $x=1$, moves toward the bulk 
of the system as  $\alpha$ is increased for a given 
$\gamma>\gamma_c$. The shock continues to exist
till the other shock phase boundary is reached.
At the shock phase boundary, the outer solution satisfying 
the boundary condition $\rho(x=1)=\gamma$ spans almost the entire lattice
except for a narrow region for the inner solution whose 
saturation value at $x=0$ just matches the 
boundary density $\alpha$.
 With further increase of $\alpha$, the shock 
disappears from the system and the system  enters
 into a different  phase where there is no shock. The density 
profile in this phase typically appears as in figure \ref{fig:highdensity}.
\highdensityplot 
Following the  case of ASEP with only mutual exclusion, 
we call this phase as 
the high-density phase. 
 In other words, as we approach 
the shock phase from  the high-density side by decreasing $\alpha$,
 the boundary layer deconfines from the $x=0$ boundary in the form of a 
shock as soon as  $\alpha$ becomes smaller than 
the saturation value of the inner solution at $x=0$.  
As a result the 
 phase boundary between the shock phase and the high-density phase is 
given by 
\begin{eqnarray}
\rho_2(\rho_{\rm o}')=\alpha,\label{shockhigh}
\end{eqnarray}
where $\rho_{\rm o}'$ is the value of the outer solution at 
the boundary $x=0$. Since the outer solution now satisfies the 
boundary condition at $x=1$, $\rho_{\rm o}'$ 
is a function of $\gamma$ and $\Omega$. 
Same equation is true for the phase boundary for $\gamma<\gamma_c$ except 
for the fact that here $\rho_{\rm o}'$ is independent of $\gamma$ since 
$\rho(x=1)=\gamma_c$.
This leads to a vertical phase boundary between the shock and 
the high-density phases. 
With the phase boundaries specified as above, the 
 phase diagram appears  as in figure \ref{fig:phasediag}.
\translines
The boundary transition exists in the high-density phase also. Depending 
on whether the value of $\alpha$ is larger or smaller than 
$\rho_{\rm o}'$, the slope of the boundary layer at $x=0$ 
changes sign. 
Since the  boundary transition lines in 
the high-density phase can be calculated following the 
principles mentioned above and also from reference \cite{smvm}, 
we skip those calculations here.

\end{document}